\documentclass[12pt]{article}

\usepackage{moreverb}
\usepackage{verbatim}
\usepackage{parskip}
\usepackage{enumerate}
\usepackage{setspace}
\usepackage{amsmath}
\usepackage{graphicx}
\usepackage{multirow}
\usepackage{colortbl}
\usepackage{float}
\usepackage{multicol}
\usepackage{subfigure}
\usepackage{enumitem}
\usepackage{bbm}
\usepackage{rotating}
\usepackage[square,numbers]{natbib}
\usepackage[left=2cm,top=2cm,right=2cm, bottom=2cm]{geometry}
\raggedright
\newcommand\BibTeX{{\rmfamily B\kern-.05em \textsc{i\kern-.025em b}\kern-.08em
T\kern-.1667em\lower.7ex\hbox{E}\kern-.125emX}}

\PassOptionsToPackage{final}{graphicx}

\begin{document}

\title{Simultaneous  Record Linkage and Causal Inference With Propensity Score Subclassification}

\author{Joan Heck Wortman\footnote{PhD candidate, Department of Statistical Science, Duke University, Durham, NC 27708 (Email:
joan.heck@duke.edu)} and Jerome P. Reiter\footnote{Professor of Statistical Science, Duke University,
Durham, NC 27708}\footnote{This research was supported by the grant NSF SES 11-31897.}}
\maketitle

\abstract{We develop methodology for causal inference in observational studies when using propensity score subclassification on data constructed with probabilistic record linkage techniques.  We focus on scenarios where covariates and binary treatment assignments are in one file and outcomes are in another file, and the goal is to estimate an additive treatment effect by merging the files. We assume that the files can be linked using variables common to both files, e.g., names or birth dates, but that links are subject to errors, e.g., due to reporting errors in the linking variables.  We develop methodology for cases where such reporting errors are independent of the other variables on the files. We describe conceptually how linkage errors can affect causal estimates in subclassification contexts. 
We also present and evaluate several algorithms for deciding which record pairs to use in estimation of causal effects.  Using simulation studies, we demonstrate that some of the procedures can result in improved accuracy in estimates of treatment effects from linked data compared to using only cases known to be true links.}


\section{Introduction}\label{sec1}

Increasingly, researchers are linking data collected in planned
studies to data available in administrative sources, such as
electronic health records  and Medicare claims data, in order to
enhance analyses of causal questions. 
For example, linking can enable researchers to evaluate long-term outcomes, as well as
outcomes not measured in the planned study, without expensive de novo
primary data collection.  It also can allow researchers to incorporate
important covariates not collected in the planned study, thereby
reducing effects of unmeasured confounding and facilitating more
nuanced estimation of treatment effects. 

When perfectly measured unique identifiers, such as
Medicare patient IDs or social security  numbers, are
available on both files, the linkage is a relatively straightforward
task: one simply merges on the identifiers and proceeds with statistical inference.  In many settings, however,
such identifiers are unavailable on at least one file, e.g., because
of privacy restrictions, and record linkage must be based on indirect
identifiers like birth dates, diagnosis codes, demographic characteristics, and names that could
differ on the files for the same individual. In such contexts, typical record linkage procedures involve scoring  potentially linked record pairs based on similarity of the linking fields---where larger values imply more confidence in the correctness of the proposed link---and selecting as links those pairs whose score exceeds some threshold. \cite{fs, jw, jin2003efficient, herzog,  christen2012data}

Methodologies for record linkage with indirect identifiers and
        for causal inference in observational data are well established;
        however, we are not aware of methodology developed 
        specifically for causal inference with linked observational data.  
Yet, the fact that we seek causal inferences clearly affects the consequences of incorrect linkages. 

 For  example, suppose that a researcher has some File A that contains  treatment  and covariate values for a set of patients, and some File B that contains long term outcomes for these and other patients.  The researcher uses propensity score matching \cite{rosenbaum1983, rosenbaum} to create balanced treatment and control groups from File A.  In this case, incorrect linkages for records excluded from the matched control set do not affect the causal estimates, whereas incorrect links for those in the treated and matched control sets do.  This example suggests general questions.  When estimating a causal effect, should we use only linked pairs where the link has 
        near certain probability of being correct, or can we benefit from
        allowing lower probability links to enter the causal
        estimate?  If the latter, how do we draw the line on what to include
        and exclude? As far as we can tell, these questions have not been addressed in the literature.

In this article, we begin to address these questions.  

 Specifically, we present and evaluate several algorithms for estimation of additive treatment effects when using subclassification on propensity scores with inexactly linked data. We consider observational studies where File A includes a binary treatment and covariate values, and File B includes outcome values.  
 We develop algorithms assuming that the processes generating mismatches in the linking variables across files are unrelated to other variables on the files; for example, errors in the names or birth dates in the two files are independent of the outcomes, treatments, and causally relevant covariates.

The basic strategy underpinning the different algorithms is as follows.  First, we order the pairs selected by the record linkage procedure from highest to lowest linking scores.  Second, we peel off the cases deemed to represent correct links with near certainty. Third, starting from these certainty cases, we sequentially concatenate new linked records to the sample previously used in treatment effect estimation, each time computing some criterion intended to increase when adding inexact matches.  Finally, we find the set of cases that corresponds to the smallest value of the criterion, and use this set of records in the causal inference.  As we demonstrate in simulation studies,  case selection procedures following this strategy can reduce mean squared errors compared to using only the certainty cases or using more liberal thresholds.

The remainder of the article is organized as follows.  In Section \ref{background}, we provide background on subclassification on propensity scores and on threshold based record linkage techniques. In Section \ref{framework}, we discuss the effects of linkage errors on causal inferences when using propensity score subclassification.  In Section \ref{finescale}, we describe several algorithms for choosing record pairs.  In Section \ref{sec:sims},  we present simulation results that compare the different algorithms and illustrate their potential benefits and limitations.  In Section \ref{conclusions}, we summarize the findings and suggest future research topics.

\section{Background} \label{background}

We require both the stable unit treatment value assumption (SUTVA) \cite{rubin1978, rubin1980randomization, rubin1990comment} and strong ignorability.  \cite{rubin1978, rosenbaum} SUTVA requires that one unit's treatment status does not affect another unit's potential outcomes and also that there are no hidden levels of treatment. 
Strong ignorability requires that all units have a non-zero probability of being in the treatment and control groups, and that treatment assignment depends only on observed covariates.

Setting aside complexities associated with three or more possible treatments, which we leave for future consideration, we focus on scenarios where there is a binary treatment. Let $w_i=1$ indicate that individual $i$ is  assigned  treatment, and  $w_i=0$ indicate that individual $i$ is  assigned control.  Let $x_i$ indicate a $p \times 1$ vector of causally relevant covariates for individual $i$.  Let $Y_i(1)$ be the value of the outcome for individual $i$ when $w_i=1$, and $Y_i(0)$ be the value of the outcome for individual $i$ when $w_i=0$.  Let $\tau_i = Y_i(1) - Y_i(0)$ be the treatment effect for individual $i$. Throughout, we assume additive treatment effects, that is,  $\tau_i = \tau$ for all individuals $i$.  Finally, let $y_i = w_iY_i(1) + (1 - w_i)Y_i(0)$ be the observed outcome for any individual $i$.

We reserve the term link (and its derivatives) for when some record in File A and some record in File B are deemed to belong to the same individual, and the term match for operations involving balancing covariate distributions in treated and control groups.

\subsection{Propensity scores and subclassification}\label{sec:pscores}

Propensity scores are used in a variety of ways in causal inference, \cite{stuartmatching, imbensrubin} including matching, inverse probability weighting, and subclassification as we do here.

The propensity score is defined as $e(x)=P(w=1|x)$, i.e., the probability of being assigned treatment given covariate pattern $x$.
It can be shown that the treatment assignment is independent of $x$ given $e(x)$.  Thus, treated and control units with the same propensity score have the same distribution of $x$, so that analysts who compare treated and control units with the same propensity score effectively remove any confounding effects from $x$ when estimating treatment effects. \cite{rosenbaum}
Given sets of individuals assigned to treatment and control, analysts can estimate each individual's  $e(x_i)$ using binary regression techniques, such as  logistic regression, where  the outcome is treatment status and the predictors are the relevant covariates.  

In propensity score subclassification, the goal is to partition the collected data into $J$ strata, called subclasses, in which treated and control units have similar covariate distributions.  The partition often is based on equally spaced quantiles of the propensity scores, e.g., every twentieth percentile.  Analysts manually adjust the breaks as necessary to ensure sufficient sample sizes in each subclass.  In the simulations, we use the common choice of $J=5$ and breaks based on manual specifications of propensity score quantiles.

Let $j \in \{1, \dots, J\}$ index the $J$ subclasses, and let $\mathcal{S}_j$ where $j=1, \dots, J$ represent the set of individuals in subclass $j$.  For each $j$, let $n_{1j}$ and  $n_{0j}$ be the number of individuals in $\mathcal{S}_j$ with $w_i=1$ and $w_i=0$, respectively.  Let 
$\bar{y}_{1j} = \sum_{i \in \mathcal{S}_j} w_i y_i/n_{1j}$ and $\bar{y}_{0j} = \sum_{i \in \mathcal{S}_j} (1-w_i) y_i/n_{0j}$. 
Within each subclass $j=1, \dots, J$, we compute the estimated subclass average treatment effect,
%\begin{eqnarray}
%\label{taujest}
$\hat{\tau}_j=\bar{y}_{1j}-\bar{y}_{0j}$.
%\end{eqnarray}
We  estimate $\tau$ using the weighted average,
\begin{eqnarray}
\label{tauest}
\hat{\tau}=\sum_{j=1}^J \lambda_j \hat{\tau_{j}}.
\end{eqnarray}
A typical value of $\lambda_j$, which we use in the simulations, is $\lambda_j = \frac{n_j}{n}$, where $n_j= n_{1j}+n_{0j}$ and $n=\sum_j n_j$.

For estimated variances, it is common to use 
\begin{eqnarray}
 \hat{var}(\hat{\tau}) = \sum_{j=1}^J \lambda_j^2 \hat{var}(\hat{\tau}_j) = \sum_{j=1}^J \lambda_j^2 \left(\frac{s_{0j}^2}{n_{0j}}+\frac{s_{1j}^2}{n_{1j}}\right), \label{varest}
\end{eqnarray}
where $s_{0j}^2 = \sum_{i \in \mathcal{S}_j} (y_i (1-w_i) - \bar{y}_{0j})^2 /(n_{0j}-1)$ and $s_{1j}^2 = \sum_{i \in \mathcal{S}_j} (y_iw_i - \bar{y}_{1j})^2 /(n_{1j}-1)$.  We note that this variance estimator is not unbiased for the true variance of \eqref{tauest}, as it does not account for estimation of the propensity scores. \cite{williamsonetal:12}

Residual imbalance often remains after subclassification. To reduce the effects of the remaining imbalance, analysts can regress $y$ on $w$ and some subset of $x$ within the subclasses.  \cite{rosenbaum, dagostino}
Let $\hat{\beta_{j}}$ be the estimated coefficient of the indicator for $w$ in the regression in subclass $j$. To estimate $\tau$, we can use
%\begin{eqnarray}
$\hat{\tau}_{\beta}=\sum_{j=1}^J \lambda_j \hat{\beta_{j}}$.
%\end{eqnarray}
We can estimate the variance using \eqref{varest}, replacing the two-sample variance estimator with the estimated variance of $\hat{\beta}_{j}$ from each within-subclass regression.

\subsection{Record Linkage}\label{sec:reclink}

We consider scenarios where an analyst seeks to link two files, File A comprising $n_A$ records and File B comprising $n_B$ records, using imperfect linking variables present in both files.   In such settings, 
many analysts use the probabilistic  record linkage framework formalized by Fellegi and Sunter.  \cite{fs}  For all possible record pairs $(i, i')$, where record $i$ is in File A and record $i'$ is in File B, the analyst computes some measure $S(\gamma_{ii'})$ that reflects the similarity of the linking variables for record $i$ from File A to those for record $i'$ from File B.  Record pairs with similarity scores above an analyst-specified threshold are declared links, and others are declared either non-links or uncertain status.  Uncertain links can be sent to clerical review for adjudication or, as is often done, treated as non-links as we do here. 

More precisely, suppose that we have $F$ linking variables.  For each field  $f \in (1, \dots, F)$, let $\gamma_{fii'}$ be a score reflecting the similarity in field $f$ for that pair. Typically, we set $\gamma_{fii'}=1$ when the values of field $f$ for records $i$ and $i'$ are identical or within some acceptable tolerance, and  set $\gamma_{fii'}=0$ otherwise.  
 For each record pair $(i, i')$, let $\gamma_{ii'}=(\gamma_{1ii'}, \dots, \gamma_{Fii'})$ be the vector comprising the comparisons for each linking field. Following Fellegi and Sunter, \cite{fs} we assume that $\gamma_{ii'}$ is a random realization from a mixture of two distributions, one for true links and one for non-links.  Let $\mathcal{M}$ be the set of true links in File A and File B, and let $\mathcal{U}$ be the set of non-links in these files. 
The mixture model for $\gamma_{ii'}$ is thus
\begin{eqnarray}
\gamma_{ii'} \mid (i,i') \in \mathcal{M} &\sim& f(\theta_m) \\
\gamma_{ii'} \mid (i,i') \in \mathcal{U} &\sim& f(\theta_u), 
\end{eqnarray}
where  $\theta_m$ and $\theta_u$ are parameters specific to each class.  For computational simplicity, usually one assumes conditional independence of the $\gamma_{fii'}$ both across fields and pairs, computing 
\begin{eqnarray}
m(\gamma_{ii'}) &=& P(\gamma_{ii'} \mid \theta_m, (i,i') \in \mathcal{M}) = \prod_f P(\gamma_{fii'}  \mid \theta_{mf}, (i,i') \in \mathcal{M}) = \prod_f \theta_{mf}^{\gamma_{fii'}}(1-\theta_{mf})^{1-\gamma_{fii'}} \\
u(\gamma_{ii'}) &=& P(\gamma_{ii'} \mid \theta_u, (i,i') \in \mathcal{U}) = \prod_f P(\gamma_{fii'}  \mid \theta_{uf}, (i,i') \in \mathcal{U}) = \prod_f \theta_{uf}^{\gamma_{fii'}}(1-\theta_{uf})^{1-\gamma_{fii'}}. 
\end{eqnarray}

Fellegi and Sunter \cite{fs} use a decision-theoretic approach to minimize Type I and Type II error rates, that is, erroneously linking or erroneously not linking records, respectively. They compute the likelihood ratio, $R(i, i') =\frac{m(\gamma_{ii'})}{u(\gamma_{ii'})}$. Values of $R(i, i')$ above some upper threshold are deemed links, and values below some lower threshold are deemed non-links. 
%If between the two thresholds, it may be submitted for review. 
When all linking fields are binary and one assumes conditional independence, it is common to write $R(i, i')$ as
\begin{eqnarray}
 \label{similarity}
 S(\gamma_{ii'})= \sum_{f=1}^F log_2\left(\frac{\theta_{mf}}{\theta_{uf}}\right)\gamma_{fii'} + log_2\left(\frac{1-\theta_{mf}}{1-\theta_{uf}}\right)(1-\gamma_{fii'}).
\end{eqnarray}
$S(\gamma_{ii'})$ is often called the linking score for pair $(i,i')$.

String data, including names, complicate the construction and computation of similarity scores. \cite{newcombe, jaro1989, jaro1995, larsen} 
A common approach, which we use here and now review briefly, is to compute Jaro-Winkler scores \cite{jw} for the string fields.

Suppose we seek to compare two strings on a set of characters, where the string in File A has $d$ such characters and the string in File B has $r$ such characters.  Suppose the two strings have $c>0$ of these characters in common and $t$ characters that are transposed. Suppose that we assign a weight to each string, say $W_A$ and $W_B$, as well as a weight to transpositions, say $W_t$.  Then, the Jaro score is 
 $\Phi_J = W_A(c/d) + W_B(c/r) + W_t(c-t)/c$.
The Jaro-Winkler score boosts the weight of agreement early in a string, resulting in $\Phi_{JW} = \Phi_J + 0.1g(1-\Phi_J)$, where $g$ is the number of characters among the first four that agree in the two strings.
Scores range from 0 (no agreement) to 1 (full agreement), and can be easily estimated using the ``jarowinkler'' function in the ``RecordLinkage'' package in R.
Analysts can use the values of $\Phi_{JW}$ as similarity measures, or turn each $\Phi_{JW}(i,i')$ into a binary $\gamma_{fii'}$ by setting $\gamma_{fii'}=1$ when  $\Phi_{JW}(i,i') > t_o$ and $\gamma_{fii'}=0$ otherwise. In the simulations of Section \ref{sec:sims}, we use this approach with $t_o = 0.95$.

The linkage process can be subject to errors, e.g., records belonging to two different individuals are linked, or incompleteness, e.g., some individuals do not appear to have links.  It is well known that incorrect and incomplete linkages can degrade
the quality of subsequent statistical inferences. \cite{herzog, gu, jaro1989,  belin}  There has been some work on accounting for such errors in inferences for regression modeling. \cite{scheuren:winkler:97, lahiri, chambers08, chambers09, chambers11, chipperfield, gutman:zasl, dalzell:reiter} We are not aware of propensity score methods for causal inference that explicitly account for inexact linkage.

\section{The effects of linkage errors on causal estimands}\label{framework}

\begin{figure}[t]
  \centering
    \includegraphics[width=15cm]{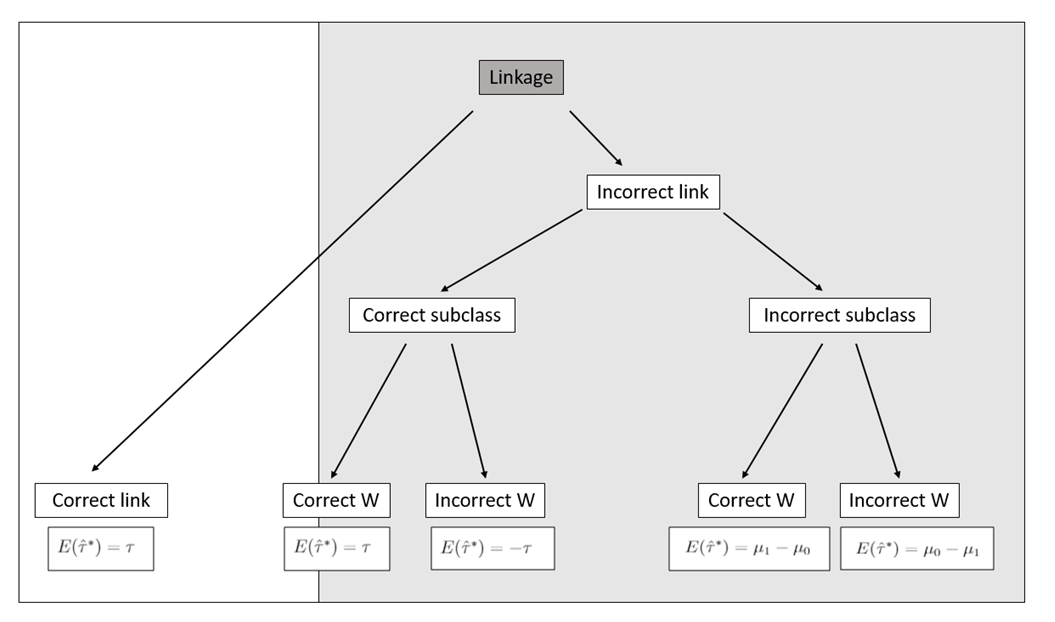}
\caption{Graphical representation of the effects of linkage errors on estimation of treatment effects in propensity score subclassification. 
$E(\hat{\tau}^*)$ is the cohort's typical value of the contribution to the expected value of the treatment effect estimator when making the corresponding linkage error. $\mu_1$ and $\mu_0$ are population marginal means of the outcomes for treated and control units, respectively.}
     \label{fig:flow}
\end{figure}

In this section, we provide intuition on the impacts of linkage errors on causal estimates made with propensity score subclassification. 
The discussion is organized around Figure \ref{fig:flow}, which highlights four types of linkage errors in the context of subclassification. Throughout we assume that it is possible to balance covariate distributions in the treatment and control groups with subclassification on properly linked records.

As we link records with $(x_i, w_i)$ measured in File A to records with $y_{i'}$ measured in File B, causal estimates are based on values that may differ from the true values due to linkage errors.  For each record $i$ in File A and its linked record $i'$ in File B, let $y_i^* = y_i$ when the linked pair is correct, i.e., records $i$ and $i'$ belong to the same individual, and let $y_i^* = y_{i'}$ when the linked pair is incorrect.  Quantities from  Section \ref{sec:pscores} use $y_i^*$ rather than $y_i$, so that, for example, the within-class estimate of treatment effect is  
\begin{eqnarray}
\label{eq:errortau}
\hat{\tau}_j^* = \bar{y}^*_{1j} - \bar{y}^*_{0j} = \sum_{i \in \mathcal{S}_j} w_i y_i^*/n_{1j}  - \sum_{i \in \mathcal{S}_j} (1-w_i) y_i^*/n_{0j}. 
\end{eqnarray}
Inferences can be based on 
\begin{eqnarray}
\hat{\tau}^* &=&  \sum_{j=1}^J \lambda_j\hat{\tau_{j}}^* \label{tauestrl}\\ 
 \hat{var}(\hat{\tau}^*) &=& \sum_{j=1}^J \lambda_j^2 \left(\frac{s_{0j}^{2*}}{n_{0j}}+\frac{s_{1j}^{2*}}{n_{1j}}\right), \label{varestrl}
\end{eqnarray}
where
$s_{0j}^2 = \sum_{i \in \mathcal{S}_j} (y_i^* (1-w_i) - \bar{y}_{0j}^*)^2 /(n_{0j}-1)$ and $s_{1j}^{2*} = \sum_{i \in \mathcal{S}_j} (y_i^*w_i - \bar{y}_{1j}^*)^2 /(n_{1j}-1)$.
As subclassification is based only on the $(x_i, w_i)$ from records in File A, values of the propensity scores are not affected by the record linkage.  Typically, the linked file used in estimation does not include all $n_A$ records from File A, in which case $\mathcal{S}_j$ should be interpreted as restricted to the set of linked pairs used in analysis, with $(n_{0j}, n_{1j})$ computed over that restricted set.

\subsection{Matching subclass and treatment status}\label{sec:MM}

It is possible to link two records that do not belong to the same individual yet not incur bias if (i) the two records' actual covariate values place them into a common subclass and (ii) they experienced the same treatment assignment.  To illustrate, suppose the record linkage algorithm links record $i$ in File A named David Copperfield  to record $i'$ in File B named Davy Copperfull. These records are not a true link, but they have similar background information. David is 43 years old, and Davy is 42. Both are men with 3 children. Both were assigned to treatment, say a new drug intended to reduce high blood pressure. If the propensity score model conditions on age, sex, and number of children, the subclassification algorithm may well put them in the same subclass $j$.

When using \eqref{tauestrl} to estimate treatment effects, linkage errors like this tend to have only modest impacts.  Suppose, for example, that this type of linkage error occurs only for two treated individuals $i$ and $k$ in subclass $j$ such that, after linkage, $y_{i}^* = y_{k} = Y_k(1)$ and $y_{k}^* = y_{i} = Y_i(1)$.  In this case, $\hat{\tau}^* = \hat{\tau}$, so there is no effect on the differences in means inference.  Of course, the regression-adjusted treatment effect estimate changes unless $x_i = x_k$.

More formally, suppose again for simplicity that only one record $k$ in subclass $j$ is incorrectly linked to a record $k'$ with the same $w$ and subclass. We consider treatment assignment within the subclass as completely randomized \cite{rosenbaum95}, that is, consider $e(x_i) = n_{1j}/n_j$ for all $i \in \mathcal{S}_j$.   In this case, averaging over the treatment assignment, we have  
\begin{eqnarray}
\label{eq:errortau1a}
\notag E(\hat{\tau}_j^*) &=& E\left(\sum_{i \in \mathcal{S}_j} w_i y_i^*/n_{1j}  - \sum_{i \in \mathcal{S}_j} (1-w_i) y_i^*/n_{0j}\right)\\
\notag &=& E\left(\sum_{i \in \mathcal{S}_j, i \neq k'} w_iy_i/n_{1j}  - \sum_{i \in \mathcal{S}_j, i \neq k'} (1-w_i) y_i/n_{0j} + w_{k'}y_{k'}/n_{1j} - (1 - w_{k'}) y_{k'}/n_{0j})\right)\\
\notag &=& \sum_{i \in \mathcal{S}_j, i \neq k'} \left(\frac{n_{1j}/n_j}{n_{1j}}Y_i(1)  - \frac{1-n_{1j}/n_j}{n_{0j}}Y_i(0)\right) + \frac{n_{1j}/n_j}{n_{1j}}Y_{k'}(1) - \frac{1-n_{1j}/n_j}{n_{0j}}Y_{k'}(0)\\  
&=& (1/n_j) \left(\sum_{i \in \mathcal{S}_j, i \neq k'} \tau_i + \tau_{k'}\right). 
\end{eqnarray}
With additive treatment effects, this expectation is still $\tau$.

An alternative way to see this is to extend to population inference. Let $(\mu_{1j}, \sigma^2_{1j})$ be the population mean and variance, respectively, of $Y_i(1)$ for all cases in subclass $j$. Let $(\mu_{0j}, \sigma^2_{0j})$ be similarly defined population quantities for all $Y_{i}(0)$ in subclass $j$.  If we think of $y_i^* = Y_{i'}(w_{i})$ as randomly drawn from the correct populations, the expectations of $\bar{y}^*_{1j}$  and $\bar{y}^*_{0j}$ continue to be $\mu_{1j}$ and $\mu_{0j}$, respectively.

\subsection{Matching subclass and non-matching treatment status}\label{sec:MN}

The record linkage algorithm might link two records with similar covariates, and hence the same subclass, that receive different treatments. This can induce substantial problems for causal estimates.  To illustrate, suppose instead that David receives the treatment but Davy does not.  We then attribute Davy's outcome to receiving the treatment rather than not receiving it.  This incorrect link biases the estimated treatment effect in the opposite direction of $\tau$. In fact, if we make this mistake many times, we could end up concluding that the treatment has the opposite effect than it truly does. 

To demonstrate this, suppose that we erroneously set a single treated individual's $y_{i}^* = Y_{k'}(0)$ where $(i, k') \in \mathcal{S}_j$.
We then have
\begin{eqnarray}
\label{eq:errortau2}
\notag E(\hat{\tau}_j^*)  &=& \sum_{i \in \mathcal{S}_j, i \neq k'} \left(\frac{n_{1j}/n_j}{n_{1j}}Y_i(1)  - \frac{1-n_{1j}/n_j}{n_{0j}}Y_i(0)\right) + \frac{n_{1j}/n_j}{n_{1j}}Y_{k'}(0) - \frac{1-n_{1j}/n_j}{n_{0j}}Y_{k'}(1)\\ 
&=& (1/n_j) \left(\sum_{i \in \mathcal{S}_j, i \neq k'} \tau_i - \tau_{k'}\right). 
%&=& (\sum_{i \in \mathcal{S}_j, i \neq k} Y_i(1) + Y_{k'}(0))/n_{1j}  - (\sum_{i \in \mathcal{S}_j, i \neq k} Y_i(0) + Y_{k'}(1))/n_{0j}. 
\end{eqnarray}

In terms of population quantities, the contribution of record $i$ to $E(\bar{y}^*_1)$ is $w_i\mu_{0j}/n_{1j}$ rather than $w_i\mu_{1j}/n_{1j}$, and the contribution to $E(\bar{y}^*_0)$ is $(1-w_i)\mu_{1j}/n_{0j}$ rather than $(1-w_i)\mu_{0j}/n_{0j}$.  The result is that $\hat{\tau}^*$ is biased.

\subsection{Non-matching subclass and matching treatment status}\label{sec:NM}

We next consider when the linking error impacts the subclass assignment but not the treatment assignment.  Suppose that 
we link record  $i$ in File A named Anna Karenina with record $i'$ in File B named Alexis Karenin.  Anna is 30 years old, female and has 2 children; Alexis is 40 years old, male and has 1 child. Neither received the treatment. When we incorrectly link Anna with Alexis, we put Alexis's $y_{i'}$ with Anna's $(x_{i}, w_{i})$. Hence, when the propensity score model is reasonable, $y_{i'}$ could be placed in an incorrect subclass, but with the correct treatment status (in this case, control). This adds bias to the treatment effect estimate.  

It is difficult to characterize the nature of this bias, since it depends on how similar the covariate distributions in the incorrectly matched subclass are to those in the actual subclass. 
What is clear is that it no longer makes sense to consider treatment assignment as completely random within the subclasses, since it is no longer reasonable to believe that covariates are balanced. 
 Hence, it is cumbersome to derive mathematical arguments like those in Sections \ref{sec:MM} and \ref{sec:MN}.  However, we can gain some insight into  this bias when we suppose that 
the process generating the linkage errors is independent of the values of $x$ and $(Y(1), Y(0))$.  In this case, we can consider the erroneous link for record $i$ to be selected randomly from cases with matching $w$    
in the incorrect subclasses.  Using the population quantities and averaging over subclasses, the contribution of record $i$ to $E(\bar{y}^*_1)$ is 
$w_i \mu_{1(-j)}/n_{1j}$, where $\mu_{1(-j)} = \sum_{h \neq j}\mu_{1h}(n_{1h}/(n_1-n_{1j}))$, rather than $\mu_{1j}/n_{1j}$. Similarly, the contribution to $E(\bar{y}^*_0)$ is $(1-w_i)
\mu_{0(-j)}/n_{0j}$ where $\mu_{0(-j)} = \sum_{h \neq j}\mu_{0h}(n_{0h}/(n_0-n_{0j}))$
rather than $\mu_{0j}/n_{0j}$.  Indeed, in an extreme case, if all records are subject to this error then one might as well not even have used subclassification, in which case $\hat{\tau}^* \approx \mu_1 - \mu_0$, 
where $\mu_1$ and $\mu_0$ are the marginal population averages of the treated and control outcomes.

\subsection{Non-matching subclass and non-matching treatment status}
Finally, we consider the case of wrong treatment and wrong subclass.  To illustrate, we link the record of an Adam Trask in file A with an Aron Trask in file B. Adam is 60 with two children, and Aron is 18 with no children. Adam has received the blood pressure medication but Aron has not. When we link Aron's outcome with Adam's background covariates and treatment indicator, we observe an incorrect link of outcome and subclass as well as an incorrect link of outcome and treatment indicator. 

As in Section \ref{sec:NM}, the potential bias induced by this type of linkage error is difficult to characterize.  When linkage errors are independent of $x$ and $(Y(1), Y(0))$, 
we can use arguments like those in  Section \ref{sec:NM}. Averaging over subclasses, the contributions of record $i$ to $E(\bar{y}^*_1)$ and $E(\bar{y}^*_0)$  are 
$w_i \mu_{0(-j)}/n_{1j}$ and $(1-w_i) \mu_{1(-j)}/n_{0j}$, respectively. 
If we make this type of mistake many times, the result will be as if we never subclassified, and we additionally labeled the treated group as the control group and vice versa, resulting in 
$\hat{\tau}^* \approx \mu_0 - \mu_1$.

\section{Stopping rules for threshold linkage}\label{finescale}

Clearly, linkage errors can have negative consequences for causal inference. One could restrict causal inference to estimating only with cases known to be true links with complete certainty. 
However, this may exclude some links that are correct, or possibly innocuously in error like those in Section \ref{sec:MM}, ultimately inflating mean squared errors.  We thus need a 
rule for deciding which linked pairs to use in $\hat{\tau}^*$.  

One approach is to attempt to choose the threshold for accepting links to minimize the mean squared error of $\hat{\tau}^*$. If we add links sequentially in decreasing order of their linkage scores, we would expect that adding the first few records to the known links should result in adding a sizable proportion of correct links.  The mean squared error  of $\hat{\tau}^*$ should decrease as we add cases to \eqref{tauestrl} until we start adding many non-links, when bias introduced by the invalid links can overwhelm the reductions in variance due to increased sample size. 

Unfortunately,  an estimator for the mean squared error of $\hat{\tau}^*$ is not apparent, as we do not know the value of $\tau$. 
Instead, we turn to a quantity that has similar behavior as the mean squared error, is easy to compute, and is familiar to users of propensity score subclassification: the estimated variance in \eqref{varestrl}.  In particular, we present three algorithms for selecting cases based on estimated variances. We derive the algorithms under the assumptions that (i) linkage errors are independent of
$x$ and $(Y(1), Y(0))$, (ii) propensity score subclassification results in groups  with balanced  covariate distributions, and (iii) treatment effects are additive. 
We assume that the analyst uses a threshold based record linkage technique like those described in Section \ref{sec:reclink}.

\subsection{The minimum estimated variance stopping rule}

The first algorithm, which we call the minimum estimated variance or MEV algorithm, is initialized as follows.  For each record pair $(i, i')$ we compute $S(\gamma_{ii'})$ using \eqref{similarity}, identifying the top match for each. When the same record from File B is 
the top match for multiple records in File A, we allow it to be used multiple times, although one could enforce one-to-one linkage.  Let $\mathcal{L}_0$ be the set of the $l_0$ record pairs known with certainty to be correct links.  We then
compute $\hat{\tau}^*$ using \eqref{tauestrl} and the estimated variance using \eqref{varestrl}, calculating the 
$\lambda$ weights and other statistics in \eqref{tauestrl} and \eqref{varestrl} from the cases in $\mathcal{L}_0$.  
We use the propensity scores and subclass boundaries determined from the analysis of all of File A.
Set a counter $h=1$.

We next arrange the top pairs in descending order of $S(\gamma_{ii'})$. Let $[h]$ index the rank order of the $h$th record pair, so that $[1]$ is the pair not in $\mathcal{L}_0$ with the highest linking score, $[2]$ is the pair not in $\mathcal{L}_0$ with the second highest linking score, and so on. 
 We append record pair $[h]$ to $\mathcal{L}_{h-1}$ to create $\mathcal{L}_{h} = \mathcal{L}_{h-1} \cup (x_{[h]}, w_{[h]}, y^*_{[h]})$.  We repeat this process for all pairs with linking scores above some minimum 
threshold, as values below this threshold are considered known not to be links, each time incrementing $h$ by one. As a result, we have a collection of $L \leq (n_A - l_0)$ successively larger 
sets $\mathcal{L}_h$.  We evaluate \eqref{tauestrl} and \eqref{varestrl} for each set of cases in $\mathcal{L}_h$, re-computing $\lambda$ each time but using the propensity scores and subclass boundaries based on all of File A. 
We select $\mathcal{L}_{min} = \{\mathcal{L}_h : h = \arg \min_{h} \hat{var}(\hat{\tau}^*)\}$. 
% We call this the minimum estimated variance stopping rule (MEV). 

As we add correct links, the estimated variance tends to decrease due to the increase in sample size.  In fact, even adding errors like those in Section \ref{sec:MM} still can result in 
decreased estimated variance, as we generally add sample size while still drawing from the correct marginal distributions of the outcomes within each subclass.  However, when we 
add pairs with other types of linkage errors, we add draws from incorrect marginal distributions, causing the estimated variance to tend to increase.  Thus, the MEV procedure tends to favor adding cases that are correct links and links with errors like those in Section \ref{sec:MM} and to disfavor adding incorrect links of other types.

To gain further insight, we present a rough approximation to the expected value of \eqref{varestrl}.  For simplicity, we  
ignore uncertainty due to estimating propensity scores and subclass boundaries, and treat observations within subclasses as independent. To motivate why the criterion is useful, suppose we consider the linked data in $\mathcal{L}_h$ as a sample from a hypothetical population of linked datasets using that threshold. 
%In other words, we view the $y_{i}^*$ as random draws from the marginal distributions of the outcomes.
For $j=1, \dots, J$, let $S^2_{h0j} = E(s_{h0j}^{2*})$ and $S^2_{h1j} = E(s_{h1j}^{2*})$, where we add the subscript $h$ to emphasize that the quantities are computed with the cases in $\mathcal{L}_h$.  For any $\mathcal{L}_h$, let 
$\mathcal{C}_{hj}$ be the set of correct links and $p_{hj}$ be the probability of a randomly sampled link within subclass $j$ being correct.      
Within any subclass $j$, we can write $S^{2}_{h0j}$ with an iterated variance,
\begin{eqnarray}
S^2_{h0j} = E(Var(y^* \mid w=0, \mathcal{C}_{hj})) + Var(E(y^* \mid w=0, \mathcal{C}_{hj})).\label{itervar}
\end{eqnarray}
Let $\mu_{h0j}$ and $\sigma^2_{h0j}$ be the  population mean and variance of $y_i^*$ for all erroneously linked records with $w_i=0$ when using the threshold associated with $\mathcal{L}_h$. For the first term of \eqref{itervar}, we have
\begin{eqnarray}
E(Var(y^* \mid w=0, \mathcal{C}_{hj})) = S^2_{h0j} p_{hj}+ \sigma^2_{h0j}(1-p_{hj}). 
\end{eqnarray}
For the second term of \eqref{itervar}, we have
\begin{eqnarray}
Var(E(y^* \mid w=0, \mathcal{C}_{hj})) 
= (\mu_{0j} - \mu_{h0j})^2 p_{hj}(1-p_{hj}).
\end{eqnarray}
We can derive a similar expression for $S^2_{h1j}$.  

Putting it all together, we have the approximation, 
\begin{eqnarray}
\notag E(\hat{var}(\hat{\tau}^*_j)) &\approx& 
p_{hj}\left(\frac{S_{0j}^2}{n_{h0j}}+\frac{S_{1j}^2}{n_{1hj}}\right)+(1-p_{hj})\left(\frac{\sigma^2_{h0j}}{n_{h0j}}+\frac{\sigma_{h1j}^2}{n_{h1j}}\right)\\
 &+& \left(\left(\frac{1}{n_{h0j}} (\mu_{0j}-\mu_{h0j})\right)^2+\left(\frac{1}{n_{h1j}} (\mu_{1j}-\mu_{h1j})\right)^2\right) p_{hj}(1-p_{hj}).\label{expectedvar}
\end{eqnarray}
The  first term in \eqref{expectedvar} is the variance for correct links within the subclass, weighted by the proportion of correct links. The second term is a variance contribution from the incorrect links. Generally, we expect the $\sigma^2_{hwj}$ to exceed the corresponding $S_{wj}^2$, since for any $(w,j)$ the distribution of $y_i^*$ for incorrect links generally should be more dispersed than the corresponding distribution of $y_i$ for  correct links, as evident from the consequences of linkage error described in Section \ref{framework}.    The third term can be viewed as a penalty for introducing incorrect links.

Using \eqref{expectedvar}, we see that \eqref{varestrl}  tends to be smallest in expectation when $p_{hj}$ is large and when linkage errors do not cause substantial differences between the means and variances of the outcomes for the correct and incorrect link cases.   Thus, using the criterion should  favor  thresholds where the fraction of true links is high and the consequences of mistakes are low, which can help improve the accuracy of treatment effect estimates compared to using only cases known to be true links.

\subsection{The estimate-tethered stopping rule}\label{extrarules}

While MEV penalizes bias as desired, it has the potential to result in undesirable case selection decisions.    
To see this, consider a scenario where the number of correct links is small compared to the number of incorrect links, and 
the treatment effect is small relative to the marginal variance of the outcome variable.  In this case, \eqref{expectedvar} could be smallest when one includes as many links as possible.  Adding cases, even incorrect links, increases the sample sizes used in \eqref{expectedvar}, which could reduce the variance terms in \eqref{expectedvar} by enough to overwhelm the increase caused by bias.

To reduce the potential for such undesirable selections, we restrict $\mathcal{L}_h$ to sets where the corresponding $\hat{\tau}^*$ is within $k$ standard errors of the estimated treatment effect based on links known to be correct.  From this set, we select the $\mathcal{L}_h$ with the minimum estimated variance.  We call this the estimate-tethered stopping rule, abbreviated as ETSR.  Formally, we choose $\mathcal{L}_{ETSR} = \{\mathcal{L}_h : h = \arg \min_{h} \hat{var}(\hat{\tau}^*), \,\, \hat{\tau}_{\mathcal{L}_h}^* \in \hat{\tau}_{\mathcal{L}_0}^* \pm  k\sqrt{\hat{var}(\hat{\tau}_{\mathcal{L}_0}^*)}\}$. 
We use $k = 0.5$ in the simulation results reported in Section \ref{sec:sims}, but the results are relatively robust to choices of $k$ between 0.5 and 2. We present results for ETSR under different values of $k$ in the supplement. Generally we expect ETSR to result in more conservative linkage than MEV, but with less room for bias.

\subsection{The minimum estimated difference-in-outcomes variance stopping rule} 

The MEV and ETSR use the propensity score subclassification when computing their respective criteria. As suggested by reviewers, some analysts may  prefer to separate the propensity score analysis from the linkage decisions as much as possible; see Section \ref{conclusions} for additional discussion of this point. We therefore propose the minimum estimated difference-in-outcomes variance rule, which we abbreviate MEDOV.  We select $\mathcal{L}_{MEDOV} = \{\mathcal{L}_h : h = \arg \min_{h} \hat{var}(\bar{y}_{1h}^* - \bar{y}_{0h}^*)\}$, 
where $\bar{y}_{1h}^*$ and  $\bar{y}_{0h}^*$ are the marginal means of the treated and control units in $\mathcal{L}_h$.   
 MEDOV is similar to MEV, but we estimate the variance before subclassification, i.e., use only $J=1$ class in \eqref{varest}.

\section{Simulation studies}\label{sec:sims}

In this section we present results of simulation studies evaluating the performance of the case selection algorithms from Section \ref{finescale}.  We base all simulations on 
the RL10000 data from the ``RecordLinkage'' package in R. \cite{rl} This dataset includes full name separated into four fields and birth dates on 9000 individuals.  For 1000 of these 9000 individuals, the RL10000 
dataset also includes duplicate records
with typographical errors on some of the fields.  No other variables are available on the file.  In all simulations, we use first name, last name, birth month, and birth day as linking variables. 

In the simulations reported here, 
we split the 10000 records into File A comprising $n_A = 2000$ records and File B comprising $n_B=8000$ records.   
In each simulation run, File A includes the 1000 records with duplicates and a random sample of 1000 records without duplicates.  File B includes the 1000 duplicates with errors and the remaining 7000 records without duplicates.
Due to the random sampling of non-duplicates across runs, the threshold values and linkage quality can change across the simulations. The effects of such changes are minor. 
%We use Fellegi-Sunter linkage in the simulations.  

For each of the 1000 records in File A with duplicates in File B, we modify the birth year of its true link in
File B so that both have the identical birth year. This allows us to block on birth year, i.e., require pairs to have the same birth year if they are to be considered links, 
and link on first name, last name, birth month, and birth day.  Blocking on birth year reduces the comparison space, which improves the quality of links and reduces computational time.  Blocking is a standard practice in record linkage settings. \cite{herzog}
We also allow units in File B to be linked to more than one unit in File A, primarily for computational convenience in repeated simulation studies. This has minimal impact on the simulation results, 
as typically only zero to two duplicates are used for the thresholds with high (greater than 95\%) link rates.  

In all simulations, we present results based on $\hat{\tau}^*$ and its variance without any regression adjustments; results for regression adjustment are in the online supplementary material.  
We estimate propensity scores using a logistic regression with treatment indicator as the outcome and main effects of the covariates as predictors. We use the full data set in File A to calculate the propensity scores.  However, in any $\mathcal{L}_h$, we use the sample sizes in the linked data  to calculate each $\lambda_j$.

As a baseline, we compare results to treatment effect estimates that would be obtained if all cases in File A were perfectly linked, i.e., all 1000 records with a link are put together and the remainder are designated non-links.  We refer to these as true links or correct links, and refer to the results as the ``Perfect'' results.  
We also compare results to the most conservative linkage strategy, in which we use only those record pairs for which $\gamma_{fii'}=1$ for all $f$.  We call these as known links or exact links, and refer to results as the ``Known'' results.

The supplement contains results of additional simulations, including a scenario where File B comprises $n_B = 2000$ records and a scenario where we use Jaro-Winkler linkage.  Results are qualitatively similar.

%%%%%%%%%%%%%%%%%%%%%%%%%%%%%%%%%%%%%%%%%%%%5

\subsection{Data generation and linkage methods}\label{sim1}

As RL10000 has no other variables, for each record $i$, we generate its treatment indicator $w_i$, two covariates $(x_{i1}, x_{i2})$, and outcome $y_i$ as follows.  We sample each $w_i$ from a 
Bernoulli distribution with probability .5.  Given $w_i$, we 
sample $x_{1i}$ from a Poisson distribution with mean $(8-3*w_i)$. We sample $x_{2i}$ from a normal distribution with mean $-w$ and standard deviation 3.  In this way, the 
distribution of $x_i$ differs for treated and control units, making propensity score subclassification useful compared to estimating $\tau$ as the difference in the marginal means.

We generate outcomes according to six different scenarios, each with an additive treatment effect.  In the first four, we use a linear response surface
\begin{equation}
y_i = 5 + 5x_{i1}+3x_{i2}+\tau w_{i}+ \epsilon_i,\,\, \epsilon_i \sim N(0, \sigma^2). \label{outcome1}
\end{equation}
In the first scenario, we set $(\tau = 50, \sigma = 10)$ to represent a treatment effect that is large relative to the variance of the outcome.
In the second scenario, we set $(\tau = 10, \sigma = 10)$  to assess the impact of having a more modest treatment effect. In the third scenario, we set $(\tau = 1, \sigma = 10)$ to examine the impact of a small, but still non-zero, treatment effect relative to the variance of the outcome.
In the fourth scenario, we set $(\tau = 50, \sigma = 25)$  to assess the impact of increasing the variance of the outcome when the treatment effect is large.
In the fifth scenario, we make the covariates have a stronger association with the outcome, using
\begin{equation}
y_i = 5 + 15x_{i1}-7x_{i2}+\tau w_{i}+ \epsilon_i,\,\, \epsilon_i \sim N(0, \sigma^2) 
\end{equation}
with $(\tau = 50, \sigma = 10)$. Finally, in the sixth scenario, we assess the impact of having a non-linear response surface, using
\begin{equation}
y_i = 5+ 0.2x_{i1}^2+exp(0.7x_{i2})+ \tau w_{i}+ \epsilon_i, \,\, \epsilon_i \sim N(0, \sigma^2)
\end{equation}
with $(\tau = 50, \sigma = 10)$.

For each simulation run, we resample values of $(w_i, x_{1i}, x_{2i}, y_i)$ for all records.  %TRUE? YES!
In all scenarios, using propensity score subclassification improves the causal estimates substantially. For example, in the first scenario where  
 $(\tau = 50, \sigma = 10)$ and the response surface is  linear, the difference in marginal means of the outcome for treated and control cases is around 31.  Using the true links for all cases in File A, $\hat{\tau}$ is around 47.  
Thus, subclassification allows for substantial bias reduction.

To implement the Fellegi-Sunter record linkage, we first block on birth year, requiring links to have the same values of birth year. We then compare the first and last names of pairs of records using Jaro-Winkler scores, which are dichotomized and fed into the  Fellegi-Sunter algorithm. For each name $f \in \{1, 2\}$, we classify $\gamma_{fii'}=1$ when $\Phi_{JW}(fii') >  .95$ and 
$\gamma_{fii'}=0$ otherwise; that is, the Jaro-Winkler score must exceed 0.95 for the fields to be called in agreement. We compare birth month and birth day using binary exact agreement indicators.
% so that $\gamma_{fii'}=1$ when $f_i = f_{i'}$ and 0 otherwise.
We compute linkage scores for each record pair that agrees on birth year using \eqref{similarity}. 
We 
set $\theta_{mf} = .95$ for all $f$ and set $\theta_{uf}$ as 
 frequency of agreement in field $f$.  
We consider record pairs with $S(\gamma_{ii'})<0$ in \eqref{similarity} not to be links.

\subsection{Results}\label{sim:results}

Table \ref{linktab2} summarizes  the quality of links at different thresholds for the first simulation scenario. Results are similar for other scenarios.  Linkage quality at thresholds of 9.3 and above is high but deteriorates quickly as one drops the threshold, with more false links and duplicates. 
The supplementary material includes examples of the linked data under different thresholds, illustrating the types of errors tolerated
when decreasing the threshold.  Of course, in applications we generally are not able to determine the link rates at different thresholds, and
hence not able to identify agreeable threshold values.  This motivates consideration of the case selection procedures.

\begin{table}[t]
\centering
\begin{tabular}{rlll}
  \hline
Threshold & Link Rate & Units & Duplicates \\ 
  \hline
0.3 & 76.2 \% & 1309 & 27 \\ 
  0.8 & 84.6 \% & 1179 & 8 \\ 
  1.6 & 91.6 \% & 1088 & 3 \\ 
  2.1 & 95 \% & 1048 & 0 \\ 
  2.8 & 97.5 \% & 1018 & 0 \\ 
  9.3 & 98.7 \% & 1002 & 0 \\ 
  9.8 & 99.2 \% & 973 & 0 \\ 
  10.5 & 99.7 \% & 775 & 0 \\ 
  11.8 & 100 \% & 638 & 0 \\ 
  19.5 & 100 \% & 531 & 0 \\ 
   \hline
\end{tabular}
\caption{Linkage summary under various thresholds. Link rate corresponds to the percentage of links that correctly correspond to the same person. Units refers to the number of cases in the linked dataset, and duplicates refers to the number of non-unique appearances of a person from File B. 
}  
\label{linktab2}
\end{table}

 \begin{figure}[t]
  \centering
    \includegraphics[width=13cm]{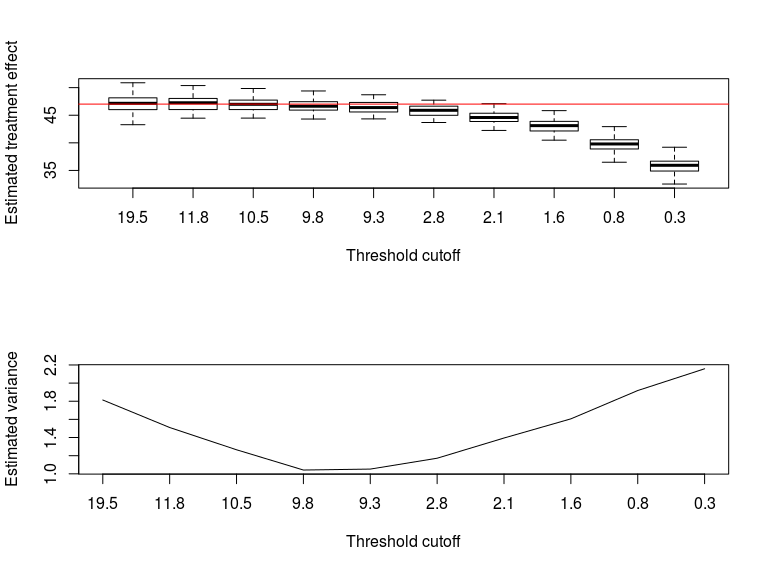}
\caption{Distribution of 100 point and variance estimates of treatment effects for simulation with Fellegi-Sunter linkage and constant treatment effect, where $\tau = 50$ and $\sigma = 10$. Horizontal line in top panel corresponds to $\hat{\tau}$ with the true links. } \label{fig1}  
\end{figure}

Figure \ref{fig1} summarizes the distributions of $\hat{\tau}^*$ and its estimated variance for 100 independent runs of the first simulation with a linear response surface and $(\tau = 50, \sigma = 10)$  at all qualifying values of the threshold for selecting cases.
The choice of threshold matters for the quality of the causal estimate.  Using thresholds below 9.3 includes incorrect links that degrade the accuracy of $\hat{\tau}^*$.  
On the other hand, using the highest threshold values cause $\hat{\tau}^*$ to be based on relatively small numbers of individuals, which results in the largest variances of $\hat{\tau}^*$.  
Apparently, the sweet spot reflecting a close-to-optimal trade off in contributions to mean-squared error is a threshold somewhere around 9.8,
which provides point estimates clustered most closely around the value of $\hat{\tau}$ attainable with the true links.  
As is evident in the bottom panel of Figure \ref{fig1}, across the 100 runs the value of \eqref{expectedvar} tends to be minimized when the threshold is around 9.8, suggesting that using the case selection algorithms could reduce mean squared errors.  

Table \ref{tab1} displays key results of the simulation runs. Turning first to the simulation setting with ($\tau = 50, \sigma = 10$), all three stopping rules reduce mean squared errors compared to using only known links.  All of the case selection methods have increased mean squared errors compared to using the true links, reflecting the information loss from having to use inexact linkage.  The percentage increases in mean squared error range from 11\% to 28\% for the case selection methods, whereas it is around 56\% for using only the known links.  ETSR offers the most substantial reductions in mean squared error, although all three methods have comparable performances.  The  average thresholds selected by MEV and ETSR are around 9.1 and 10.5, respectively.  MEDOV is more conservative, in that it adds the smallest number of links to the known links.

Turning to the second and third scenarios where we reduce $\tau$, the three case selection methods offer substantial reductions in mean squared errors compared to using only the known links.  
The reductions in mean squared error for MEDOV are not as substantial as those for MEV and ETSR, mostly because it does not add many links to the known cases as evident in the large average thresholds.  
In the scenario with $\tau = 1$, interestingly, MEV has a smaller mean squared error than using all the true links.  This results partly because $\tau$ is close to zero. In this scenario, MEV ends up using some incorrect links that bias the treatment effect toward zero, which is close enough to $\tau$ to reduce mean squared error compared to using all the true links. 

In the fourth scenario where we return $\tau = 50$ and increase the variance to $\sigma=25$, we again see that the case selection methods have larger mean squared errors than using the true links, as one would expect generally.   Here, however, the MEV has a somewhat larger bias, which happens because the MEV criterion accepts too many false links, as evident by the average threshold level of 6.7.  Because of the large variance in the outcomes, accepting false links tends to have greater impact on the bias and mean squared error in $\hat{\tau}^*$. This illustrates the concerns about MEV noted at the beginning of Section \ref{extrarules} and used to motivate ETSR.  In contrast, ETSR and MEDOV continue to substantially outperform using the known links only, with ETSR offering slightly larger reductions than MEDOV.

In the fifth scenario, the stronger associations between the covariates and the outcome increase variances, and hence mean squared errors, of the treatment effect estimators compared to the previous scenarios. Here, ETSR has the smallest mean squared error among the case selection procedures. In contrast, MEDOV performs worse than using the known cases alone.

Finally, in the sixth scenario where the response surface is non-linear, all three case selection procedures are again preferable to using known links alone.  Here ETSR offers the greatest reductions in mean squared error, getting to almost the same mean squared error as using the true links. Evidently, tethering the estimates helps ensure that low quality links are not added to the sample used for the treatment effect estimate.

\begin{table}[t]
\centering
\singlespacing
\begin{tabular}{crrrrrr}
  \hline
Scenario & Linkage & Mean $\hat{\tau}^*$ &  Var ($\hat{\tau}^*$) & $\hat{var}(\hat{\tau}^*)$  & MSE $\hat{\tau}^*$ & Threshold\\
   \hline
      \multirow{6}{6cm}{Linear, $(\tau = 50, \sigma = 10)$} & 
 Perfect & 47.0 & 1.0 & 0.9 & 9.9 &  \\ 
&  Known & 47.0 & 6.8 & 6.9 & 15.4 &  \\ 
 &  MEV & 46.6 & 1.1 & 1.0 & 12.7 & 9.1 \\ 
  & \textbf{ETSR} & \textbf{47.0} & \textbf{1.9 }& \textbf{1.2} & \textbf{11.0 }&\textbf{ 10.5} \\ 
  & MEDOV & 46.9 & 1.7 & 1.4 & 11.4 & 13.1 \\ 
 \hline
   \multirow{6}{6cm}{Linear, $(\tau = 10, \sigma = 10)$} & 
Perfect & 7.1 & 1.1 & 0.9 & 9.3 &  \\ 
&  Known & 7.3 & 7.2 & 6.5 & 14.3 &  \\ 
&  MEV & 7.0 & 1.1 & 1.0 & 9.9 & 7.5 \\ 
&  \textbf{ETSR} & \textbf{7.1} & \textbf{1.7} & \textbf{1.1} & \textbf{9.9} & \textbf{8.2} \\ 
&  MEDOV & 6.8 & 1.7 & 1.3 & 12.1 & 11.1 \\ 
  
   \hline
    \multirow{6}{6cm}{Linear, $(\tau = 1, \sigma = 10)$}
    
& Perfect & -1.9 & 1.1 & 0.9 & 9.3 & \\ 
 & Known & -1.7 & 7.2 & 6.5 & 14.3 &  \\ 
 & \textbf{MEV} &\textbf{ -1.8} & \textbf{1.1} & \textbf{1.0} & \textbf{8.9} & \textbf{7.6} \\ 
 & ETSR & -1.8 & 1.7 & 1.2 & 9.8 & 9.1 \\ 
 & MEDOV & -2.1 & 1.6 & 1.4 & 10.9 & 12.7 \\ 
  
  \hline
  \multirow{6}{6cm}{Linear, $(\tau = 50, \sigma = 25)$} & 
 Perfect & 47.3 & 4.8 & 4.0 & 12.1 &  \\ 
&  Known & 47.7 & 33.7 & 27.4 & 38.7 & \\ 
&  MEV & 46.0 & 10.5 & 4.1 & 26.6 & 6.7 \\ 
&  \textbf{ETSR} & \textbf{47.0} & \textbf{8.1} &\textbf{ 4.8} & \textbf{17.1} & \textbf{8.8} \\ 
&  MEDOV & 46.9 & 7.4 & 5.9 & 17.2 & 13.8 \\ 
 \hline
  \multirow{6}{6cm}{Linear, High $R^2$, $(\tau = 50, \sigma = 10)$} 
& Perfect & 43.9 & 6.5 & 7.0 & 43.7 &  \\ 
 &  Known & 44.5 & 48.9 & 46.9 & 79.0 &  \\ 
  & MEV & 42.5 & 13.4 & 7.1 & 69.3 & 6.1 \\ 
  & \textbf{ETSR} & \textbf{43.8} & \textbf{12.9} & \textbf{8.1} & \textbf{51.5} & \textbf{7.9} \\ 
  & MEDOV & 41.7 & 21.6 & 9.5 & 90.8 & 9.8 \\ 
  \hline
 \multirow{6}{6cm}{Non-linear, $(\tau = 50, \sigma = 10)$} & 
Perfect & 48.1 & 22.7 & 25.8 & 26.3 &  \\ 
&   Known & 48.1 & 96.3 & 104.7 & 99.0 &  \\ 
 &  MEV & 43.3 & 29.2 & 5.8 & 74.4 & 5.8 \\ 
 &  \textbf{ETSR} & \textbf{47.0} & \textbf{17.8} & \textbf{9.5} & \textbf{26.7} &\textbf{ 9.8} \\ 
  & MEDOV & 45.6 & 29.6 & 7.2 & 48.8 & 9.8 \\ 
  \hline
\end{tabular}
\caption{Summary of results across 100 runs of the six simulation scenarios with Fellegi-Sunter linkage. Var($\hat{\tau}^*)$ refers to the empirical variance of the estimated treatment effects across each set of 100 runs, and $\hat{var}(\hat{\tau}^*)$ refers to the average of the estimated variances across each set of 100 runs. Threshold refers to the average value of the threshold chosen across the 100 runs.  ``Perfect" refers to the analysis using all 1000 true links.  ``Known"  refers to the analysis with all records in File A with $\gamma_{fii'}=1$ for all $f$. Results with the lowest MSE for each simulation are in bold. \label{tab1}}  
\end{table}

\section{Concluding Remarks}\label{conclusions}

Methods for causal inference and record linkage have developed independently, but the simulation results indicate that it can be fruitful to consider methods that explicitly account for both tasks. 
For settings where covariates and assignments are in one file and outcomes are in the other file, 
the simulations here and in the supplementary material suggest that case selection strategies can improve causal estimates for analyses based on propensity score subclassification. In these simulations, arguably ETSR  performs best overall. It has the smallest mean squared errors in some scenarios, and when other stopping rules have lower mean squared errors, ETSR generally is not far behind. 
Of course, these findings are based on limited simulation studies and particular assumptions, most importantly that the linkage errors are approximately independent of $(x, w, Y(1), Y(0))$.  
Future research is needed to assess the performance of the case selection strategies when this independence assumption is not reasonable.

As with any methodology, there are scenarios where the case selection procedures may not be effective.  In particular, when the known links include outliers that pull the estimate 
of $\hat{\tau}_{\mathcal{L}_0}$ far away from $\tau$, the procedures might not add many cases, even true links, to the data used in the causal estimate, as doing so could cause the estimated variance to increase.
Additionally, the case selection procedures may not be relevant when one seeks to estimate non-additive treatment effects.  Finally, the procedures may suffer when the underlying analysis models are poorly specified,  
including the propensity score models, subclass boundaries, and regressions for adjusted inferences.

The case selection procedures are designed to work with common record linkage techniques like the Fellegi-Sunter approach. A potential alternative is to adapt record linkage techniques that 
average over different compositions of the linked population.  For example, Gutman et al. \cite{gutman:zasl} and Dalzell and Reiter \cite{dalzell:reiter} 
sample from the posterior distribution of a  
latent linking matrix, informed by a posited regression model that connects an outcome variable in File A to predictors in File B.  Adapting such approaches specifically for causal inference 
is an intriguing area for future research.

We recommend being sensible in the choice of linkage technology. We found that
 adding many incorrect links, e.g., by using very low thresholds to accept almost any proposed link, can reduce the estimated variance of the treatment effect due to the 
increased sample size. However, this results in poor quality estimates of $\tau$.  When reasonable cutoffs on linkage scores are enforced, 
%as is nearly always done in practice,
 the pattern of the estimated variance under varying thresholds tends to be U-shaped. However, the estimated variance can become S-shaped when large numbers of incorrect links are added. 
Using the ETSR limits the possibility of favoring thresholds corresponding to high numbers of incorrect links, but we still emphasize the importance of using sound record linkage techniques when making causal inferences with linked data.

Finally, we close with a comment on the philosophy of causal inference in observational studies.  Many researchers follow the guidance to separate the design of the study from 
the analysis. \cite{imbensrubin}  The case selection procedures partially adhere to that guidance. When the covariates and treatment are in the same file, one can estimate propensity
scores and form subclasses without referring to the outcomes.  However, the procedures utilize the outcomes when selecting the sample to use for estimation.  
If one seeks the potential gains in accuracy from adding more links, this is the price to pay for working with imperfect data.

\end{document}

\title{Simultaneous Record Linkage and Causal Inference with Propensity Score Subclassification- Online Supplement}
\author{Joan Heck Wortman and Jerome P. Reiter}
%\date{2016}

\maketitle

\section{Introduction}
In this supplement, we present additional simulation results, organized as follows. In Section \ref{regressioncorrection}, we show results from simulation studies using the same data generation methods as the main text, but using a regression correction within subclasses as part of the causal analysis. In Section \ref{tether}, we examine the simulation results from the main text simulations using the ETSR with values of $k$ ranging from 0.1 to 3. In Section \ref{extrasims}, we present simulation results for two additional scenarios: one using a smaller size of File B and one using average Jaro-Winkler scores as the record linkage technique. 

\section{Original simulation studies with regression correction}\label{regressioncorrection}

In this section we present results of simulation studies evaluating the performance of the case selection algorithms outlined in Section 4 of the main text with an additional within-subclass regression correction in the causal analysis.  We use the same data generation process as in Section 5.1 of the main text. 

The only difference in the analysis is that within each subclass, we add a regression correction.   according to the model $y^* = x_{1}^*\alpha_{1j} + x_2^*\alpha_{2j} + w^*\beta_j + \epsilon_{j}$, where $\epsilon_j \sim N(0, \sigma^2)$ and estimating all parameters via the usual ordinary least squares. We use the same model for each simulation, including the final simulation with a non-linear response surface. Let $\hat{\beta_{j}}$ be the estimated coefficient of the indicator for $w$ in the regression in subclass $j$. To estimate $\tau$, we can use
%\begin{eqnarray}
$\hat{\tau}_{\beta}=\sum_{j=1}^J \lambda_j \hat{\beta_{j}}$.
%\end{eqnarray}
We can estimate the variance using (2) in Section 2.1 of the main text, replacing the two-sample variance estimator with the estimated variance of $\hat{\beta}_{j}$ from each within-subclass regression.

Here we repeat Table \ref{linktab2} from the main text, so as to provide context for the linkage quality.  As noted in the main text, the quality of links at thresholds of 9.8 and above is high but deteriorates quickly as one drops the threshold, with more false links and duplicates. Table \ref{exampletab1} shows examples of the linked data under different thresholds, illustrating the types of errors tolerated
when decreasing the threshold.  

\begin{table}[t]
\centering
\begin{tabular}{rlll}
  \hline
Threshold & Link Rate & Units & Duplicates \\ 
  \hline
0.3 & 76.2 \% & 1309 & 27 \\ 
  0.8 & 84.6 \% & 1179 & 8 \\ 
  1.6 & 91.6 \% & 1088 & 3 \\ 
  2.1 & 95 \% & 1048 & 0 \\ 
  2.8 & 97.5 \% & 1018 & 0 \\ 
  9.3 & 98.7 \% & 1002 & 0 \\ 
  9.8 & 99.2 \% & 973 & 0 \\ 
  10.5 & 99.7 \% & 775 & 0 \\ 
  11.8 & 100 \% & 638 & 0 \\ 
  19.5 & 100 \% & 531 & 0 \\ 
   \hline
\end{tabular}
\caption{Linkage summary under various thresholds. Link rate corresponds to the percentage of links that correctly correspond to the same person. Units refers to the number of cases in the linked dataset, and duplicates refers to the number of non-unique appearances of a person from File B. }  
\label{linktab2}
\end{table}

\begin{table}[t]
\scriptsize
\centering
\begin{tabular}{rllllrrrrrrl}
  \hline
Threshold & F. Name:A & F. Name:B & L. Name:A & L. Name:B & Mo:A & D:A & Y:A & Mo:B & D:B & Y:B & Status \\ 
  \hline
0.30 & FRIEDA & GERHARD & MUELLER & MUELLER &   8 &  25 & 1941 &   8 &  14 & 1941 & False \\ 
  0.80 & RENATE & RENATE & SCHMIDT & WERNER &  11 &  12 & 1939 &  11 &  19 & 1939 & False \\ 
  1.61 & PETRA & KLAUS & SCHMITT & SCHMITT &   7 &  14 & 1958 &   6 &  14 & 1958 & False \\ 
  2.10 & KLAUS & KLAUS & WAGNER & KUEHN &   5 &  14 & 1968 &   9 &  14 & 1968 & False \\ 
  2.83 & HEINZ & HEINZ & MAYER & MAYER &   7 &  13 & 1949 &  12 &   2 & 1949 & False \\ 
  9.26 & PAUL & PAFUL & PFEIFFER & PFEIFFER &  10 &  20 & 1956 &  10 &  20 & 1956 & True \\ 
  9.76 & CHRISTINE & CHRISTINE & MUELLER & MUEKLER &   7 &  18 & 1937 &   7 &  18 & 1937 & True \\ 
  10.49 & BAERBEL & BAERBEL & FISCHER & FISCHER &   8 &   7 & 1976 &   8 &   4 & 1976 & True \\ 
  11.80 & FRANK & FRANK & PETERS & PETERS &   6 &   1 & 1990 &   7 &   1 & 1990 & True \\ 
  19.45 & GERTRUD & GERTRUCD & MUELLER & MUELLER &  11 &  19 & 1986 &  11 &  19 & 1986 & True \\ 
   \hline
\end{tabular}
\caption{Examples of new links added under various thresholds. Field values are shown for first name, last name, birth day, birth month, and birth year in files A and B, along with true (but unobserved) link status.}  
\label{exampletab1}
\end{table}

 \begin{figure}[t]
  \centering
    \includegraphics[width=15cm]{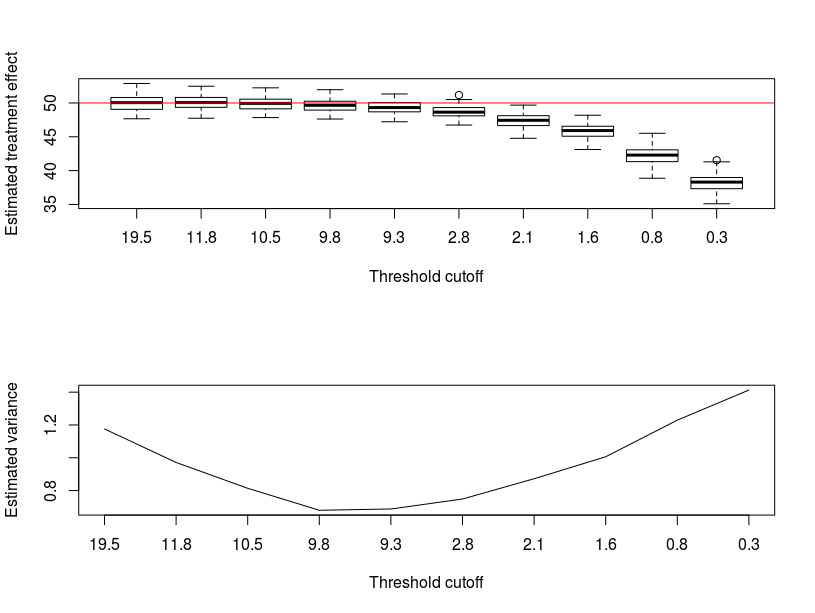}
\caption{Distribution of 100 point and variance estimates of treatment effects for simulation with Fellegi-Sunter linkage and constant treatment effect, where $\tau = 50$ and $\sigma = 10$. Horizontal line in top panel corresponds to $\hat{\tau}$ with the true links.} \label{fig1}  
\end{figure}

% \subsubsection{Subclassification only}
Figure \ref{fig1} summarizes the distributions of $\hat{\tau}^*$ and its estimated variance for 100 independent runs of the first simulation with a linear response surface and $(\tau = 50, \sigma = 10)$  at all qualifying values of the threshold for selecting cases, illustrating the behavior of the estimates under various thresholds.
Similar to analysis without a regression correction, the choice of threshold matters for the quality of the causal estimate.  Using thresholds below 9.8 includes incorrect links that degrade the accuracy of $\hat{\tau}^*$.  
On the other hand, using the highest threshold values cause $\hat{\tau}^*$ to be based on relatively small numbers of individuals, 
which results in relatively high variances of $\hat{\tau}^*$.

Table \ref{tab1} displays key results of the simulation runs. First examining the simulation setting with ($\tau = 50, \sigma = 10$), all three stopping rules reduce mean squared errors compared to using only known links.  All of the case selection methods have increased mean squared errors compared to using the true links, reflecting the information loss from having to use inexact linkage. In this simulation, MEV offers the most substantial reductions in mean squared error, although all three methods have comparable performances.  The next two simulations with $\tau = 10$ and $\tau = 1$ have qualitatively similar results.

In the fourth scenario where we return $\tau = 50$ and increase the variance to $\sigma=25$, we again see that the case selection methods have larger mean squared errors than using the true links, as one would expect generally.   Here, however, the MEV does not perform as well as using only the known cases. With MEV, the estimated variance of $\hat{\tau}^*$ is substantially smaller than its true variance, which happens because the MEV criterion accepts too many false links, as evident by the average threshold level of 3.5.  Because of the large variance in the outcomes, accepting false links tends to have greater impact on the bias and mean squared error in $\hat{\tau}^*$. As a result, MEV increases bias and decreases the accuracy of the variance
estimation.  This illustrates the concerns about MEV noted at the beginning of Section 4.2 of the main text and used to motivate ETSR.  In contrast, ETSR and MEDOV continue to outperform using the known links only, with MEDOV offering slightly larger reductions than ETSR.

In the fifth scenario, 
MEV has the smallest mean squared error among the case-selection procedures, though ETSR is not far behind. In contrast, MEDOV has the worst performance, driven by both the higher bias and the higher variance. 
MEDOV results in large variation in the selected thresholds.  The simulated standard error of the threshold choice for MEDOV across the 100 runs of this simulation is around 6.6, compared to 0.7 for MEV.

In the sixth scenario where the response surface is non-linear, all three case selection procedures are again preferable to using known links alone.  Here ETSR offers the greatest reductions in mean squared error.

\begin{table}[t]
\centering
\singlespacing
\begin{tabular}{crrrrrr}
  \hline
Scenario & Linkage & Mean $\hat{\tau}^*$ &  Var ($\hat{\tau}^*$) & $\hat{var}(\hat{\tau}^*)$  & MSE $\hat{\tau}^*$ & Avg. $h$\\
   \hline
      \multirow{6}{7cm}{Linear, $(\tau = 50, \sigma = 10)$} & 
Perfect & 50.0 & 0.6 & 0.1 & 0.6 &  \\ 
&  Known & 49.9 & 4.9 & 4.5 & 4.9 &  \\ 
&  MEV & 49.6 & 0.9 & 0.7 & 1.1 & 9.6 \\ 
&  ETSR & 49.9 & 1.4 & 0.8 & 1.3 & 10.8 \\ 
&  MEDOV & 49.8 & 1.2 & 0.9 & 1.2 & 13.1 \\ 
 \hline
\multirow{6}{7cm}{Linear, $(\tau = 10, \sigma = 10)$} &  
Perfect & 10.1 & 0.7 & 0.1 & 0.7 &  \\ 
&   Known & 10.2 & 5.7 & 4.4 & 5.7 &  \\ 
 &  MEV & 10.0 & 0.8 & 0.6 & 0.8 & 9.3 \\ 
  & ETSR & 10.1 & 1.3 & 0.8 & 1.2 & 9.2 \\ 
  & MEDOV & 9.7 & 1.2 & 0.9 & 1.3 & 11.1 \\
  \hline

    \multirow{6}{7cm}{Linear, $(\tau = 1, \sigma = 10)$}
    
& Perfect & 1.1 & 0.7 & 0.1 & 0.7 &  \\ 
 &  Known & 1.2 & 5.7 & 4.4 & 5.7 &  \\ 
 &  MEV & 1.1 & 0.8 & 0.6 & 0.8 & 9.3 \\ 
 &  ETSR & 1.1 & 1.2 & 0.8 & 1.2 & 10.2 \\ 
 &  MEDOV & 0.9 & 1.1 & 0.9 & 1.1 & 12.7 \\ 
  \hline
  \multirow{6}{7cm}{Linear, $(\tau = 50, \sigma = 25)$}  
& Perfect & 50.2 & 4.3 & 0.8 & 4.3 &  \\ 
&   Known & 50.6 & 35.4 & 27.7 & 35.5 &  \\ 
&   MEV & 44.3 & 32.0 & 3.8 & 63.8 & 3.5 \\ 
&   ETSR & 50.0 & 7.7 & 4.4 & 7.6 & 8.7 \\ 
&   MEDOV & 49.9 & 7.1 & 5.6 & 7.1 & 13.8 \\
 \hline
  \multirow{6}{7cm}{Linear, High $R^2$, $(\tau = 50, \sigma = 10)$} & 
Perfect & 50.1 & 0.7 & 0.1 & 0.7 &  \\ 
&  Known & 50.2 & 5.5 & 4.4 & 5.5 &  \\ 
&  MEV & 49.8 & 0.9 & 0.8 & 1.0 & 9.9 \\ 
&  ETSR & 50.1 & 1.3 & 0.9 & 1.3 & 11.7 \\ 
&  MEDOV & 47.8 & 19.8 & 1.6 & 24.5 & 9.8 \\ 
 \hline
 \multirow{6}{7cm}{Non-linear, $(\tau = 50, \sigma = 10)$} & 
 Perfect & 50.9 & 16.9 & 3.5 & 17.6 & \\ 
 &  Known & 51.0 & 98.1 & 79.7 & 98.2 &  \\ 
 &  MEV & 45.5 & 31.7 & 7.1 & 51.9 & 5.5 \\ 
 &  ETSR & 49.7 & 15.7 & 9.3 & 15.7 & 9.5 \\ 
 &  MEDOV & 47.9 & 25.4 & 7.6 & 29.6 & 9.8 \\ 
  \hline
\end{tabular}
\caption{Summary of results across 100 runs of the six simulation scenarios with Fellegi-Sunter linkage and a regression correction within subclasses. Var($\hat{\tau}^*)$ refers to the empirical variance of the estimated treatment effects across each set of 100 runs, and $\hat{var}(\hat{\tau}^*)$ refers to the average of the estimated variances across each set of 100 runs. Avg. $h$ refers to the average threshold chosen across the 100 runs.  ``Perfect" refers to the analysis using all 1000 true links.  ``Known"  refers to the analysis with all records in File A with $\gamma_{fii'}=1$ for all $f$.  \label{tab1}}  
\end{table}

\section{ETSR tether parameter}\label{tether}

In this section we present results of the simulation scenarios in Section 5 of the main text and in Section \ref{regressioncorrection} of this supplement using the ETSR with different values of $k$. We use the same methods of analysis as the main text: Fellegi-Sunter record linkage and subclassification on propensity scores without a regression correction.

Overall, the results with a tether parameter $k$ between 0.5 and 2 are qualitatively similar. Sometimes, however, setting $k = 0.1$ results in too conservative of a threshold choice, and setting $k = 3$ results in too liberal of a threshold choice.  Although $k=0.5$ is a reasonable compromise, analysts may want to choose $k$ based on the context of the analysis and the relative consequences of including too many false links or not enough true links.

\begin{table}[t]
\centering
\singlespacing
\begin{tabular}{crrrrrr}
  \hline
Scenario & $k$ & Mean $\hat{\tau}^*$ &  Var ($\hat{\tau}^*$) & $\hat{var}(\hat{\tau}^*)$  & MSE $\hat{\tau}^*$ & Avg. $h$\\
   \hline
      \multirow{6}{6cm}{Linear, $(\tau = 50, \sigma = 10)$} 
      
& 0.1 & 47.1 & 2.1 & 1.6 & 10.7 &  \\ 
  & 0.5 & 47.0 & 1.9 & 1.2 & 11.0 &  \\ 
  & 1 & 46.7 & 1.2 & 1.1 & 11.8 & 9.3 \\ 
  & 2 & 46.6 & 1.1 & 1.0 & 12.6 & 9.1 \\ 
  & 3 & 46.6 & 1.1 & 1.0 & 12.7 & 9.1 \\  
 \hline
 
  \multirow{6}{6cm}{Linear, $(\tau = 10, \sigma = 10)$}  
 & 0.1 & 7.1 & 2.1 & 1.6 & 10.5 &  \\ 
   & 0.5 & 7.1 & 1.7 & 1.1 & 9.9 &  \\ 
   & 1 & 7.1 & 1.2 & 1.0 & 9.7 & 7.9 \\ 
   & 2 & 7.0 & 1.1 & 1.0 & 9.9 & 7.5 \\ 
   & 3 & 7.0 & 1.1 & 1.0 & 9.9 & 7.5 \\ 
   \hline

    \multirow{6}{6cm}{Linear, $(\tau = 1, \sigma = 10)$}
    
 & 0.1 & -1.9 & 2.0 & 1.5 & 10.5 &  \\ 
   & 0.5 & -1.8 & 1.7 & 1.2 & 9.8 &  \\ 
   & 1 & -1.8 & 1.2 & 1.0 & 9.2 & 7.6 \\ 
   & 2 & -1.8 & 1.1 & 1.0 & 8.9 & 7.6 \\ 
   & 3 & -1.8 & 1.1 & 1.0 & 8.9 & 7.6 \\ 

  \hline
  \multirow{6}{6cm}{Linear, $(\tau = 50, \sigma = 25)$}  
 & 0.1 & 47.2 & 9.4 & 6.3 & 17.0 &  \\ 
   & 0.5 & 47.0 & 8.1 & 4.8 & 17.1 &  \\ 
   & 1 & 46.7 & 6.3 & 4.3 & 17.0 & 7.7 \\ 
   & 2 & 46.4 & 5.9 & 4.1 & 19.0 & 7.1 \\ 
   & 3 & 46.2 & 7.7 & 4.1 & 22.3 & 6.9 \\ 
 \hline
  \multirow{6}{6cm}{Linear, High $R^2$, $(\tau = 50, \sigma = 10)$} 
& 0.1 & 44.0 & 14.2 & 10.7 & 50.4 & \\ 
  & 0.5 & 43.8 & 12.9 & 8.1 & 51.5 & \\ 
  & 1 & 43.5 & 10.6 & 7.4 & 53.1 & 7.1 \\ 
  & 2 & 43.1 & 8.3 & 7.1 & 56.3 & 6.5 \\ 
  & 3 & 42.6 & 12.5 & 7.1 & 66.9 & 6.2 \\ 
 \hline
 \multirow{6}{6cm}{Non-linear, $(\tau = 50, \sigma = 10)$} 
 & 0.1 & 47.7 & 21.9 & 14.6 & 26.9 &  \\ 
   & 0.5 & 47.0 & 17.8 & 9.5 & 26.7 &  \\ 
   & 1 & 46.0 & 15.2 & 6.5 & 30.9 & 7.9 \\ 
   & 2 & 44.6 & 20.4 & 6.0 & 49.4 & 6.2 \\ 
   & 3 & 43.8 & 26.1 & 5.8 & 64.6 & 5.9 \\ 
   \hline
\end{tabular}
\caption{Summary of results across 100 runs of the six simulation scenarios with Fellegi-Sunter linkage and propensity score subclassification. $k$ refers to the number of standard errors used in the tether restriction. Var($\hat{\tau}^*)$ refers to the empirical variance of the estimated treatment effects across each set of 100 runs, and $\hat{var}(\hat{\tau}^*)$ refers to the average of the estimated variances across each set of 100 runs. Avg. $h$ refers to the average threshold chosen across the 100 runs. \label{tethertab}}  
\end{table}

\section{Additional simulation studies}\label{extrasims}
In this section, we present results from two additional simulations: one with a smaller size of File B and one using average Jaro-Winkler scores as the record linkage method.  

\subsection{Simulation with smaller size for File B}

In the first simulation, we examine the performance of the algorithm when the two files to be linked are the same sizes. Instead of linking File A to a File B with 8,000 records, we leave only 2,000 records in File B, half of which have a link in File A. 
We use the data generation process and record linkage techniques in the first scenario in Section 5 of the main text. i.e., the response surface is linear with $(\tau = 50, \sigma = 10)$, and generate 100 independent simulation runs.

Table \ref{linktab8} shows the match rate, number of linked units, and the number of File B duplicate units under the different possible thresholds. The match rates here are higher than in the other simulations, since there are fewer false links possible with the smaller size of File B. Therefore, even at the lowest threshold above zero, we still see a match rate of 97.9 \%. Table \ref{extab8} shows randomly selected examples of links added under the various thresholds. Although the thresholds are numerically similar to the other simulations, the implications on linkage quality are clear. The pool of possible links is a higher quality due to the smaller size of File B, so the links added are more likely to correspond to the same person.

\begin{table}[t]
\centering
\begin{tabular}{llll}
  \hline
Threshold & Link Rate & Units & Duplicates \\ 
  \hline
0.6 & 97.9 \% & 1020 & 0 \\ 
  0.8 & 98.7 \% & 1011 & 0 \\ 
  1.3 & 99.8 \% & 996 & 0 \\ 
  8 & 99.9 \% & 991 & 0 \\ 
  8.3 & 99.9 \% & 967 & 0 \\ 
  8.8 & 100 \% & 773 & 0 \\ 
  9.8 & 100 \% & 638 & 0 \\ 
  17.3 & 100 \% & 531 & 0 \\ 
   \hline
\end{tabular}
\caption{Linkage summary under various thresholds for simulation with smaller File B size. Link Rate corresponds to the percent of links that correctly correspond to the same person. Units refers to the size of the linked data set, and duplicates refers to the number of non-unique appearances of a person from File B.}  
\label{linktab8}
\end{table}

\begin{table}[ht]
\scriptsize
\centering
\begin{tabular}{rllllrrrrrrl}
  \hline
Threshold & F. Name:A & F. Name:B & L. Name:A & L. Name:B & Mo:A & D:A & Y:A & Mo:B & D:B & Y:B & Status \\ 
  \hline
0.57 & JUERGEN & ANGELIKA & SCHULZ & SCHULZ &   7 &  24 & 1947 &   6 &  24 & 1947 & False \\ 
  0.82 & WALTER & WALTER & KOEHLER & MEYER &   7 &  18 & 1935 &   9 &  18 & 1935 & False \\ 
  1.33 & ROBERT & ROBERT & LANG & LANG &   9 &  24 & 2007 &   2 &  22 & 2007 & True \\ 
  8.05 & ELDKE & ELKE & WEISS & WEISS &   4 &  30 & 1978 &   4 &  30 & 1978 & True \\ 
  8.30 & KARIN & KARIN & MUELLRR & MUELLER &   2 &   9 & 1974 &   2 &   9 & 1974 & True \\ 
  8.80 & RENATE & RENATE & HORN & HORN &  11 &  18 & 1994 &  11 &  81 & 1994 & True \\ 
  9.85 & RUTH & RUTH & MEIER & MEIER &  11 &  29 & 1961 &   1 &  29 & 1961 & True \\ 
  17.32 & STEFEAN & STEFAN & STEIN & STEIN &  11 &  21 & 1938 &  11 &  21 & 1938 & True \\ 
   \hline
\end{tabular}
\caption{Examples of new links added under various thresholds for simulation with smaller File B size. Field values are shown for first name, last name, birth day, birth month, and birth year in files A and B, along with true (but unobserved) link status.}  
\label{extab8}
\end{table}

\begin{figure}[H]
  \centering
    \includegraphics[width=13cm]{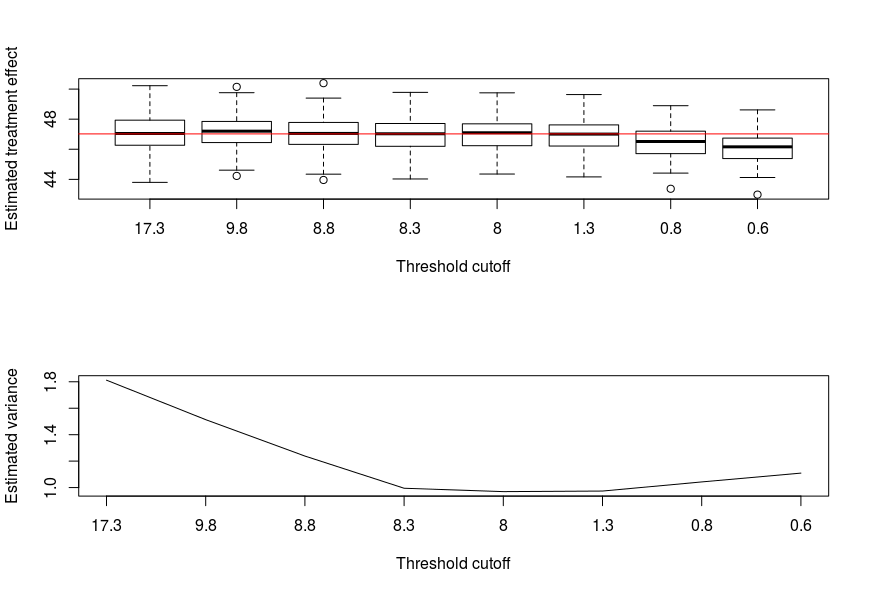}
\caption{Distribution of 100 point and variance estimates of treatment effects for simulation with Fellegi-Sunter linkage and constant treatment effect with smaller File B. Horizontal line in top panel corresponds to $\hat{\tau}$ with the perfectly linked data.} \label{fig14}  
\end{figure}

Figure \ref{fig14} and the top panel of Table \ref{newsimstab} summarize the results. All the case selection procedures offer improvements over using the known links alone, with the best performance using ETSR.

\begin{table}[t]
\centering
\singlespacing
\begin{tabular}{crrrrrr}
  \hline
Scenario & Linkage & Mean $\hat{\tau}^*$ &  Var ($\hat{\tau}^*$) & $\hat{var}(\hat{\tau}^*)$  & MSE $\hat{\tau}^*$ & Avg. $h$\\
   \hline
      \multirow{6}{7cm}{Linear, $n_B = 2000, (\tau = 50, \sigma = 10)$} 
      
& Perfect & 47.0 & 1.0 & 0.9 & 9.9 & \\ 
 & Known & 46.9 & 6.2 & 6.3 & 16.1 &  \\ 
 & MEV & 46.9 & 1.0 & 1.0 & 10.4 & 3.9 \\ 
 & \textbf{ETSR} & \textbf{47.0} & \textbf{1.3} & \textbf{1.1} & \textbf{10.0} & \textbf{5.8} \\ 
 & MEDOV & 46.9 & 1.2 & 1.4 & 11.1 & 10.3 \\ 
 \hline

    \multirow{6}{7cm}{Linear, JW Linkage, $(\tau = 50, \sigma = 10)$}
    
  & Perfect & 47.1 & 1.1 & 0.9 & 9.3 & \\ 
  & Known & 47.3 & 7.2 & 6.5 & 14.3 & \\ 
  & \textbf{MEV} & \textbf{47.0} & \textbf{1.2} & \textbf{1.1} & \textbf{10.4} & \textbf{0.9} \\ 
  & ETSR & 47.1 & 2.1 & 1.3 & 10.6 & 0.9 \\ 
 &  MEDOV & 47.0 & 2.0 & 1.5 & 11.2 & 1.0 \\ 

 \hline
\end{tabular}
\caption{Summary of results across 100 runs of the two simulation scenarios with Fellegi-Sunter or Jaro-Winkler linkage and propensity score subclassification. Var($\hat{\tau}^*)$ refers to the empirical variance of the estimated treatment effects across each set of 100 runs, and $\hat{var}(\hat{\tau}^*)$ refers to the average of the estimated variances across each set of 100 runs. Avg. $h$ refers to the average threshold chosen across the 100 runs. \label{newsimstab}}  
\end{table}

\subsection{Simulation with Jaro-Winkler scores}
The second set of simulations uses average Jaro-Winkler scores for record linkage and an additive (constant) treatment effect. We generate the data in the same way as the first simulation in Section 5 of the main text, but change the record linkage method. 

We use the similarity metric $S(\gamma_{ii'}) = (1/4) \sum_{f=1}^4 \Phi_{JW}(fii')$, where $\Phi_{JW}(fii')$ is the Jaro-Winkler similarity 
of the comparison field $f$ for record pair $(i,i')$. We require $S(\gamma_{ii'}) \geq 0.8$ for the pair to be considered a possible link.  As evident in Table \ref{linktab1}, the quality of links at thresholds of 0.9 and above is high, with a match rate upwards of 99\%, but it deteriorates at a threshold of 0.8. Table \ref{extab1} shows examples of new links under various thresholds.

\begin{table}[t]
\centering
\begin{tabular}{llll}
  \hline
Threshold & Link Rate & Units & Duplicates \\ 
  \hline
0.8 & 88.5 \% & 956 & 0 \\ 
  0.9 & 99.5 \% & 850 & 0 \\ 
  0.9 & 99.8 \% & 834 & 0 \\ 
  1 & 100 \% & 712 & 0 \\ 
  1 & 100 \% & 612 & 0 \\ 
  1 & 100 \% & 493 & 0 \\ 
   \hline
\end{tabular}
\caption{Linkage summary under various thresholds for Jaro-Winkler linkage simulation. Link rate corresponds to the percentage of links that correctly correspond to the same person. Units refers to the number of cases in the linked dataset, and duplicates refers to the number of non-unique appearances of a person from File B.}  
\label{linktab1}
\end{table}

\begin{table}[t]
\scriptsize
\centering
\begin{tabular}{rllllrrrrrrl}
  \hline
Threshold & F. Name:A & F. Name:B & L. Name:A & L. Name:B & Mo:A & D:A & Y:A & Mo:B & D:B & Y:B & Status \\ 
  \hline
0.80 & KARIN & ULRIKE & SCHNEIDER & SCHNEIDER &   9 &  12 & 1954 &   9 &  22 & 1954 & False \\ 
  0.90 & SABINE & SABINE & FRANK & FRANK &   4 &  22 & 1931 &   4 &  72 & 1931 & True \\ 
  0.92 & STEFAN & STEFAN & WAGNER & WAGNER &   8 &  21 & 1926 &   8 &  22 & 1926 & True \\ 
  0.98 & JUERGEN & JUERGEN & MURLLER & MUELLER &   4 &  14 & 2002 &   4 &  14 & 2002 & True \\ 
  0.98 & WOLFGANG & WOLFGANG & FISCHWR & FISCHER &   2 &  26 & 1967 &   2 &  26 & 1967 & True \\ 
  0.99 & NORBERT & NORBERT & KAISER & KAISER &   4 &  11 & 1934 &   4 &  11 & 1934 & True \\ 
   \hline
\end{tabular}
\caption{Examples of new links added under various thresholds. Field values are shown for first name, last name, birth day, birth month, and birth year in files A and B, along with true (but unobserved) link status.}  
\label{extab1}
\end{table}

  \begin{figure}[t]
  \centering
    \includegraphics[width=13cm]{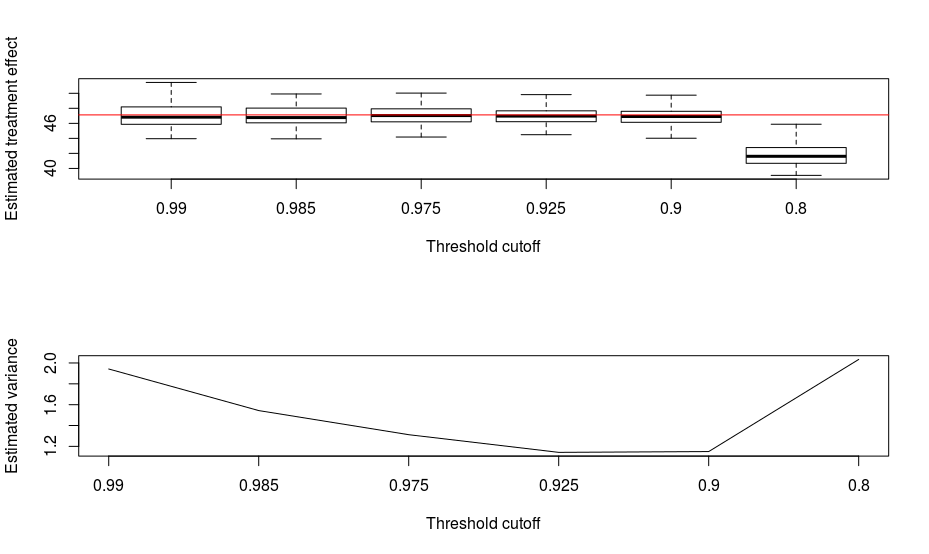}
\caption{Distribution of 100 point and variance estimates of treatment effects for simulation with Jaro-Winkler score linkage and constant treatment effect. Horizontal line in top panel corresponds to $\hat{\tau}$ with the true links.} \label{fig2}  
\end{figure}

% \subsubsection{Subclassification only}
Figure \ref{fig2} summarizes the distributions of $\hat{\tau}^*$ and its estimated variance for 100 independent runs of the simulation that uses the linear outcome distribution described for the first simulation in Section 5 of the main text with $\tau = 50, \sigma  = 10)$. Once again, the choice of threshold matters for the quality of the causal estimate, although all are reasonably high quality. 
The sweet spot reflecting a close-to-optimal tradeoff in contributions to mean-squared error is a threshold somewhere around 0.9, which provides point estimates clustered most closely around the value of $\hat{\tau}$ attainable with perfect record linkage.  
As is evident in the bottom panel of Figure \ref{fig2}, across the 100 runs the estimate of the variance tends to be minimized when the threshold is around 0.9.

Table \ref{newsimstab} summarizes results of treatment effect estimation when using the different rules for selecting links. Compared to using only the known links with all fields having $\gamma_{fii'}=1$, using any of the proposed threshold rules reduces the mean squared error of $\hat{\tau}^*$ to the point where results are similar to those based on the true links. MEV and ETSR tend to result in smaller mean square errors than MEDOV, although the difference is minor.